\documentclass[letterpaper,10pt]{article} 

\usepackage{opticameet3} 
\usepackage{mathtools}
\usepackage{siunitx}
\usepackage{cite}
\usepackage{dirtytalk}
\usepackage{ulem}

\newcommand\authormark[1]{\textsuperscript{#1}}

\newcommand{\beginsupplement}{%
        \setcounter{table}{0}
        \renewcommand{\thetable}{S\arabic{table}}%
        \setcounter{figure}{0}
        \renewcommand{\thefigure}{S\arabic{figure}}%
     }

\usepackage{amsmath,amssymb}
\usepackage{soul}
\usepackage[colorlinks=true,bookmarks=false,citecolor=blue,urlcolor=blue]{hyperref} 

\graphicspath{{Figures/}{Figures/peter/}}

\begin{document}

\title{Inverse Design of Unitary Transmission Matrices in Silicon Photonic Coupled Waveguide Arrays using a Neural Adjoint Model}

\author{Thomas W. Radford\authormark{1*}, Peter R. Wiecha\authormark{2}, Alberto Politi\authormark{1}, Ioannis Zeimpekis\authormark{3}\authormark{4}, and Otto L. Muskens \authormark{1}}

\address{\authormark{1} School of Physics and Astronomy, University of Southampton, Southampton, SO17 1BJ, United Kingdom\\
\authormark{2} LAAS, Université de Toulouse, CNRS, Toulouse, France\\
\authormark{3} School of Electronics and Computer Science, University of Southampton, Southampton, SO17 1BJ, United Kingdom\\
\authormark{4} Optoelectronics Research Centre, University of Southampton, Southampton, SO17 1BJ, United Kingdom}
\email{\authormark{*}T.Radford@soton.ac.uk} 

\begin{abstract}
The development of low-loss reconfigurable integrated optical devices enables further research into  technologies including photonic signal processing, analogue quantum computing, and optical neural networks. Here, we introduce digital patterning of coupled waveguide arrays as a platform capable of implementing unitary matrix operations. Determining the required device geometry for a specific optical output is computationally challenging and requires a robust and versatile inverse design protocol. In this work we present an approach using high speed neural network surrogate based gradient optimization, capable of predicting patterns of refractive index perturbations based on switching of the ultra-low loss chalcogenide phase change material, antimony tri-selinide ($\text{Sb}_{2}\text{Se}_{3}$). Results for a $3\hspace{-0.04cm}\times\hspace{-0.04cm}3$ silicon waveguide array are presented, demonstrating control of both amplitude and phase for each transmission matrix element. Network performance is studied using neural network optimization tools such as dataset augmentation and supplementation with random noise, resulting in an average fidelity of 0.94 for unitary matrix targets. Our results show that coupled waveguide arrays with perturbation patterns offer new routes for achieving programmable integrated photonics with a reduced footprint compared to conventional interferometer-mesh technology.

\end{abstract}

\section{Introduction}
Integrated photonics offers a platform for the miniaturization of optical devices and systems, yielding increased stability, greatly reduced size and complexity compared to traditional optical systems. Integrated photonic technologies have been successfully demonstrated across a wide range of domains including vector-matrix multiplication \cite{zhou2022photonic,giamougiannis2023neuromorphic}, quantum simulation \cite{aspuru2012photonic, wang2020integrated}, signal processing \cite{choutagunta2019adapting} and the development of high speed neural networks \cite{shastri2021photonics}. While application-specific photonic integrated circuits require a bespoke design-fabrication cycle for each variation in functionality, there is an emerging interest in platforms that can be re-programmed after fabrication to provide fine-tuning, diversification, or entirely new methods of deployment \cite{Bogaerts2020}. Re-programmable devices may also have the ability to dynamically tailor their optical output \cite{Bogaerts2020b}, and use cases were realized across a number of fields including post fabrication device processing \cite{chen2017post}, optical switches \cite{zheng2018gst} and optical signal compensation \cite{staffoli2023equalization}. One field in particular looking to utilize reconfigurable technologies is photonic computing \cite{shen2017deep, perez2017multipurpose, shastri2021photonics, ashtiani2022chip} where the development of next generation optical devices rely on the construction of structures with rapidly configurable output characteristics allowing for high speed information processing.

Commonly used reconfigurable technologies such as micro-heaters \cite{wu2016reconfigurable, atabaki2013sub} and micro electro-mechanical devices (MEMS) \cite{errando2015low, oshita2020reconfigurable, Quack2023} either introduce additional complex fabrication steps or rely on the coupling of external regulating electronics onto the photonic chip. Such approaches occupy valuable chip space and impose power constraints, while simultaneously introducing undesirable heat and noise into experimental systems. An alternative method for the production of reconfigurable devices involves using non-volatile optical phase change materials (PCMs) such as antimony tri-selenide ($\text{Sb}_{2}\text{Se}_{3}$) \cite{delaney2020new}. PCM based approaches can offer ultra-low loss, compact and passive programming capabilities, while maintaining compatibility with a wide range of optical devices across telecommunications wavelengths \cite{Miller2018OpticalReview, ChenACSPhoton2022, WuSciAdv2024, PrabhatanIScience2023, Tripathi2023}. PCMs such as $\text{Sb}_{2}\text{Se}_{3}$ are able to exist in multiple non-volatile solid states which can be programmed using optical or electrical pulses. The dielectric functions of amorphous and crystalline states of PCMs differ significantly from each other, which depending on the nature of the chemical bonds can result in large order changes in refractive index \cite{wuttig2017phase}. An approach using direct optical writing allows the addition of reconfigurability to previously fabricated devices using minimal overheads in fabrication and regulation \cite{DelaneySciAdv2021}. Endurance in programmable devices using PCM technologies is seeing improvements \cite{Martin-Monier:22} with repeated refractive index modulation of $>\hspace{-0.04cm}10^{6}$ switching cycles recently being demonstrated in thin-films of $\text{Sb}_{2}\text{Se}_{3}$ \cite{lawson2024optical, Alam2024}.

The relationship between input and output modes of any linear optical element can be described by a complex transmission matrix. Through careful design of multi-port devices, we are therefore able to produce an analogue for any given n$\times$m matrix so long as a fabrication process can accommodate its production. Integrated approaches using continuously coupled devices offer an interesting platform whereby a device can be specifically designed to implement any arbitrary transmission matrix, confined within the bounds of a micron-scale structure \cite{Miller2013}. Interference based devices with the addition of individual scattering sites are a common platform for the implementation of arbitrary optical transmission matrices  \cite{liu2011design, lu2013nanophotonic, Bruck2016, frellsen2016topology,dinsdale2021deep,nikkhah2024inverse}, as well as devices consisting of reconfigurable interferometer meshes \cite{shen2017deep, perez2017multipurpose, taballione20198, Bogaerts2020b}. 

In this work, we explore an alternative approach, based on coupled waveguide arrays (CWG) covered with a thin film of PCM programmed with a nano-scale perturbation pattern \cite{Lahini2008,Peruzzo2012, Garanovich2012,Lahini2018,Petrovic2015,Petrovic2023,Yang2024}. With focused laser writing, it is possible to reversibly modulate the local refractive index of individual pixels on the device surface with diffraction limited spatial resolution \cite{WuSciAdv2024, DelaneySciAdv2021}. This pattern facilitates modulation of the coupling coefficients between neighbouring waveguides, allowing full control over the devices transmission properties. As light propagates through a part of the device containing a switched pixel, it develops a phase shift relative to light traveling in an unswitched region due to the local effective index contrast.
Using a large enough number of these small effects at each pixel, enables full control over both the waveguide coupling behavior, as well as the phase of the light at output ports across a full 2$\pi$ range.

In contrast to micro-heater or interferometer mesh based devices, the perturbation pixel patterns in our CWG platform are correlated to the device's transmission matrix in a highly complex manner. A crucial step to develop this technology must therefore be the development of a powerful inverse design protocol, used to find a pixel pattern that implements a specific, complex transmission matrix. 
Significant research has been undertaken to develop a variety of inverse design processes. Modern techniques can be split into two overarching methods; inverse design through topology optimization \cite{bendsoe1989optimal} and the utilization of deep-neural networks. Topology optimizations are very powerful \cite{jensen2011topology} and have found applications across a wide range of disciplines, from aerospace engineering \cite{martinelli2013computational, seabra2016selective} and medical research \cite{wang2016topological, al2020structural}, to photonics \cite{lu2014topological, piggott2015inverse, lin2016cavity, sell2017large, christiansen2020fullwave}. Such optimizations typically result in complex-shaped structures with fine features which can prove difficult to fabricate, in some cases becoming unrealistic to produce at all \cite{augenstein2020inverse}.
Gradient based topology approaches often face practical limitations as gradient optimizations require smooth variables, but their design problem is typically categorized into a binary basis, where material either is present or absent. 
It should be noted that while there do exist non-gradient based optimization methods, many of these have been demonstrated as intractable for real world practical systems with large degrees of freedom \cite{sigmund2011usefulness, bennetIllustratedTutorialGlobal2024}. The most popular current alternative is the integration of artificial intelligence and machine learning into the device design pipeline, with increasing attention being paid to the creation of an AI-topological hybrid approach \cite{hegde2019photonics, ren2021genetic, woldseth2022use}. 

In this work, we demonstrate an inverse design tool chain using a gradient based optimization with a deep learning surrogate forward model \cite{peurifoy2018nanophotonic}, as well as a geometry re-parametrisation using a Wasserstein generative adversarial network (WGAN). Using this high speed technique we build on previous works in which similar methods are employed for the design and development of individual device elements, such as the height, orientation and shape of elliptic meta atom pillars \cite{deng2021neural, ren2021benchmarking}, as well as in more complex free-space design tasks \cite{augenstein2023neural}. 

\section{Neural adjoint inverse design approach}

\subsection{Device Model}

To demonstrate the capabilities of our inverse design method, we consider the case of an array of three coupled silicon rib waveguides, implementing an arbitrary $3\times3$ transmission matrix. In our simulations, these waveguides are covered by crystalline $\text{Sb}_{2}\text{Se}_{3}$, which may then be switched locally, for example experimentally by using direct laser writing \cite{DelaneySciAdv2021}. Experimentally it is easier to produce a narrow amorphization laser profile, meaning greater spatial precision can be achieved by using pixel maps with a crystalline background. We consider thin films of PCM, which have been shown experimentally as a feasible geometry \cite{DelaneySciAdv2021}, allowing a consistent and repeatable index modulation to be applied.

\begin{figure}[t!]
    \centering
    \includegraphics[width = \linewidth]{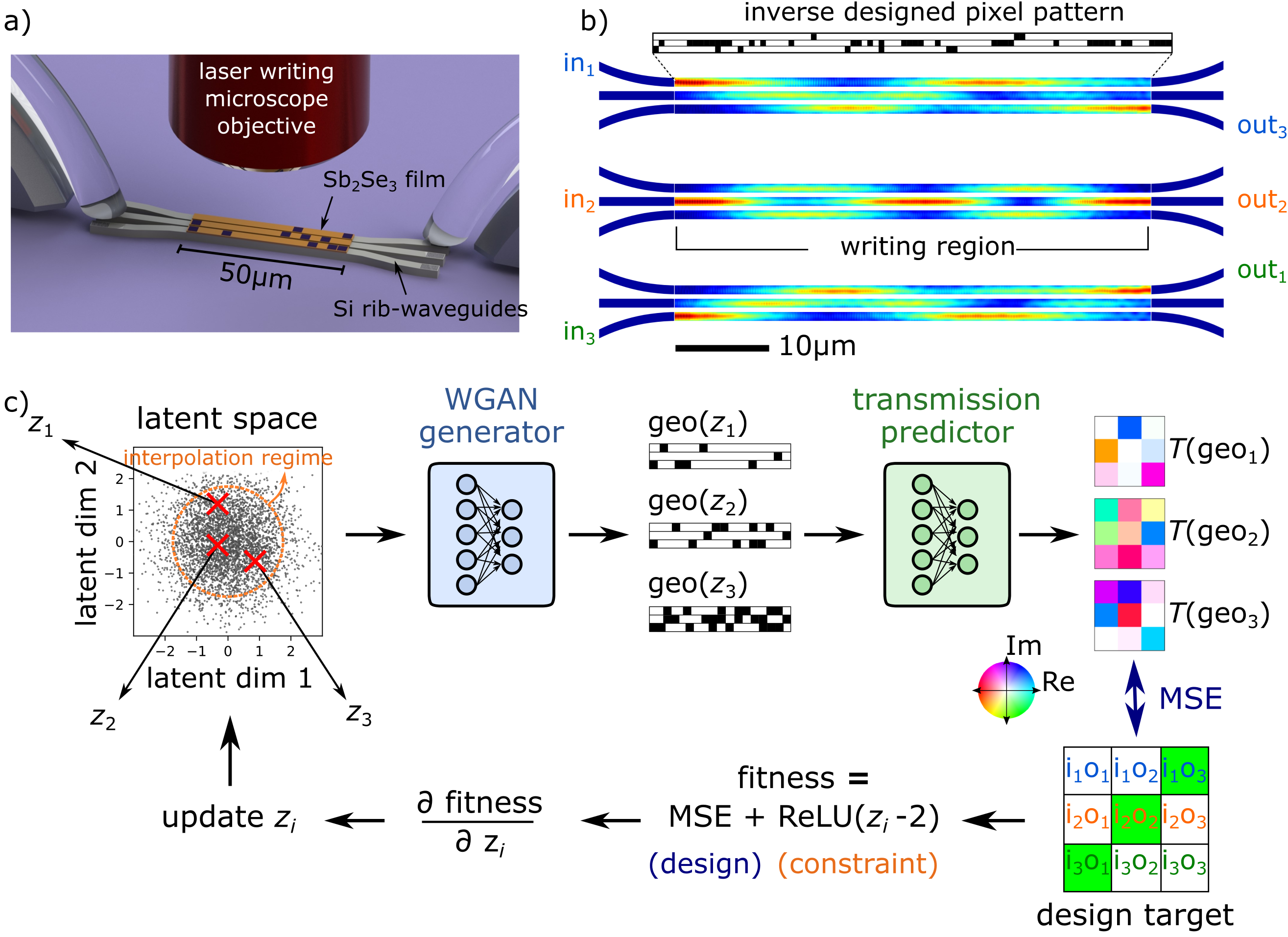}
    \caption{%
    a) Artistic view of the reconfigurable coupled waveguide device. An array of coupled silicon rib waveguides (here three) is covered by a thin PCM layer (here $\text{Sb}_{2}\text{Se}_{3}$), which can be locally switched between crystalline and amorphous state using a laser writing setup. This is parametrized in a pixel pattern, in this figure 500\,nm large pixels along 50\,\textmu m of the waveguides. The input and output ports of the waveguide array can be individually accessed by optical gratings.
    b) varFDTD simulations of the electric field intensity maps along the coupled waveguide array for three different input ports. The pattern is designed to correspond to an anti-diagonal exchange matrix with equal phases for each output.
    c) Full schematic of the inverse design pipeline, used to find the optimal pattern shown in b). 
    }
    \label{fig:cwg_and_network_scheme}
\end{figure}

Our coupled waveguide device model is based on a standard silicon on insulator wafer structure, a schematic of the device is shown in Figure\,\ref{fig:cwg_and_network_scheme}a. Structural features are etched 120\,nm deep into a 200\,nm thick silicon overlayer, which lies on top of a 2\textmu m thick buried oxide layer.
The rib waveguides are 500\,nm wide, allowing only propagation of the fundamental mode at the considered vacuum wavelength of 1550\,nm. The distance between waveguides is 250\,nm over a 50\textmu m long coupled region. To avoid any cross-talk outside of this coupling region, waveguides fan out to a spacing of 1\textmu m at both  input and output sides.
The waveguides are covered with a 30\,nm thick $\text{Sb}_{2}\text{Se}_{3}$ layer to maintain their single mode performance, displaying an index contrast between crystalline and amorphous states as high as $\Delta n_{\text{$\text{Sb}_{2}\text{Se}_{3}$}} = 0.77$. Finally, the structures are capped with a semi-infinite cladding layer of SiO$_{2}$.  
Pixels are defined on a 500\,nm square grid over the coupling region's area, consisting of a total of 96 pixels for each of the three waveguides (288 pixels in total). 

Coupled waveguide devices are simulated using the finite difference time domain (FDTD) software Lumerical, specifically we use the the Variational FDTD (varFTDT) engine, MODE. This engine collapses a 3D device structure into an effective 2D simulation by probing the effective index experienced in slices across the simulation mesh. The resultant simulation achieves comparable accuracy to a full 3D simulation, however requires significantly less computational resources. As this method accounts for the 3D structure of the device model it is commonly referred to as a 2.5D simulation. Examples are shown throughout this work, for example in figure~\ref{fig:cwg_and_network_scheme}b, showing the electric field intensity of a device with a transmission matrix analogous to the inverse identity matrix. After simulation, the devices transmission matrix can be extracted by sweeping across input waveguides, for all input modes the complex field at the peak amplitude of each output is recorded and constructed into a complex matrix. 

\subsection{Inverse design goals and training data generation}

The choice of training data is a particular important aspect of the model under study. Similar to previous studies \cite{dinsdale2021deep}, we use pre-conditioned data by specifying a target output profile for a single input port.
It is not constructive to use a training set of purely random patterns \cite{dinsdale2021deep}, because random pixel patterns typically do not co-operate in providing an effective state conversion between input and output, as discussed in more detail in Section~\ref{sec:random}. For an acceptable performance of the deep learning based design method, a more directed approach to training data generation is required, one that from the start aims towards the later inverse design task.

To this end, we first implement a brute force iterative optimization algorithm in the training data generation. The process starts by randomly choosing a specific input channel, as well as a random splitting ratio for the output channels. Then, the state of a single, random pixel is switched and the device transmission matrix is evaluated through varFDTD simulations. If, compared to the previous device geometry, the result is closer to the target splitting ratio, the pixel is retained and further simulations are run for each input channel in order to record the full transmission matrix. If not, the pixel is reverted to its initial state, another random pixel is selected, and the process is repeated. After a specified number of iterative cycles the process starts over with a blank device and a new optimization target. At each evaluation step where a perturbation is kept, the transmission of the other input ports is also evaluated and the entire pattern and its associated complex transmission matrix are saved to the dataset. In this way the dataset is populated with a large variety of pixel numbers and sufficient variance in the device performance to build a balanced dataset. A detailed description of this process as well as a specific example can be found in the supplementary information. 

Training data generation was carried out in this manner to create an initial dataset containing 21,406 pixel patterns alongside their associated complex transmission matrices. The horizontal plane of symmetry found in the unperturbed waveguide array was furthermore exploited so as to double the size of this training set. In this step we flip both the pixel pattern and transmission matrix along the midpoint of the array, allowing us to extend the training data to contain 42,812 patterns without the need for extra simulation time.

Once a database of pre-optimized pixel patterns and their associated transmission matrices is created, we repeat this process to create a separate validation dataset used to verify that the network is sufficiently generalized and may accurately predict transmission matrices for patterns from outside of the dataset it was trained on. This validation set contains a further 3282 patterns which were in turn doubled using the same method as discussed above. 

\subsection{Gradient Based Inverse Design using a Deep Learning Surrogate}


The forward predictor used is a ResNeXt encoder-decoder convolutional neural network \cite{xie2017aggregated}, the training of which took 1h using a Nvidia RTX3070 GPU. A detailed schematic of this forward network model can be found in the supplemental information figure~\ref{fig:forward_NN_sketch}. After training, the forward network is capable of predicting the complex transmission matrix for pixel patterns similar to the training geometries. However, pixel patterns that differ substantially from the training set samples (e.g. many more pixels or strongly clustered pixels), are predicted at very low accuracy.



Our predictor model is trained only on the pre-simulated dataset and has no direct knowledge of the physics in the system. While such purely data-based surrogate models are known to work accurately in the interpolation regime, it is also well known that extrapolation to geometries that differ from those in the dataset is usually very error-prone \cite{wiechaDeepLearningMeets2020, khaireh-waliehNewcomerGuideDeep2023}.
In consequence, optimizing directly the pixel patterns using the forward model as a differentiable physics predictor, fails in a remarkably reproducible manner. The gradient based optimization routinely converges towards geometries for which the forward model's predictions are wrong. In other words, optimising the pixel pattern using the neural network's design gradients, leads to geometries with unsatisfactory transmission matrices after varFDTD simulations, but, according to the forward network, should be well optimized.

Thus, a mechanism is necessary to constrain the allowed geometries during optimization to pixel patterns which the forward model is capable of faithfully predicting. Typically this is achieved through the introduction of a boundary loss term in the design optimization \cite{deng2021neural}. However, in free-form optimizations such a constraint is difficult to formulate.

\subsection*{Regularization of the Gradient Based Optimization} To avoid the optimization converging to the extrapolation regime of the forward network, we use a further deep-learning based approach. Integrating a Wasserstein generative adversarial network with gradient penalty \cite{gulrajani2017improved} (WGAN-GP), we develop a learned re-parametrization of the pixel patterns geometry from the training dataset. 
Details of the WGAN-GP architecture are shown in the supplemental information figure~\ref{fig:WGAN_sketch}. 

The key idea is, that during training, the WGAN-GP develops a mapping of the pixel pattern geometries into a compact and continuous latent space, in which the pattern geometries of the training data are normally distributed around the mean value $\mu_z=0$ and with a variance of $\sigma_z=1$. The latent vector representations of the geometries in the dataset being normally distributed in the latent space means, that it is possible to interpolate between two latent vectors. Furthermore, every intermediate point in the latent space also corresponds to a pixel pattern that is valid within the distribution of the geometries in the training set. 

In other words, the pixel-based representations of the geometries that form a non-convex set in the original parametrization (by pixels; interpolation leads to non-physical patterns with gray-scale pixels), are mapped into a convex set of latent representations of the same geometries. Any interpolation between these latent vectors should still represent a valid perturbation pattern, which can be generated from its latent vector using the WGAN-GP generator network. 
This also means, that due to the training procedure using normally distributed random sampling, all geometries that are valid interpolations of the dataset samples, lie within said normal distribution (with known $\mu_z=0$, $\sigma_z=1$). 
Therefore, to constrain the inverse design optimization loop to the interpolation regime of the neural network, we have to reformulate the design problem to find an optimum latent vector $z$ instead of an optimum pixel pattern. If we optimize in the WGAN-GP latent space, we can then simply add a regularization term to the inverse design fitness function (we use mean square error between target and predicted transmission matrix). Since the statistical properties of the WGAN-GP's latent space are known (assuming successful training of the latter), this regularization term has to penalize solutions with latent values far from the mean of the normal distribution.
Practically, we add a simple rectified linear unit (ReLU) constraint term to the design fitness function $\sum_i \text{ReLU}(z_i - 2)$, that penalizes values of $z_i$ outside of a $2\sigma$ range (see figure~\ref{fig:cwg_and_network_scheme}c).  

\subsection*{Local Minima during Gradient Descent} Gradient based optimizations such as used here are prone to become stuck in local minima, and therefore require good initial guesses to reach a globally optimum solution. To ensure that we find a close to ideal solution, we run a large number of concurrent optimizations for each design target (100 in our case). Eventually, the best ranked solution is kept. We note that due to the highly optimized GPU-based parallelization of modern deep learning frameworks such as tensorflow used in this work, the concurrent optimization of many targets is very efficient and fast.

\subsection*{Inverse Design Loop} A full schematic of our inverse design pipeline is shown in figure \ref{fig:cwg_and_network_scheme}c. A set of several latent vectors is randomly initialized. The WGAN generator (blue) predicts the corresponding geometries, of which the complex transmission matrices are predicted by the forward network (green). The latter are compared to the design target transmission using the mean square error as a metric. The total fitness comprises an additional constraint term to restrict the optimization to a $2\sigma$ region of the WGAN latent space, which should roughly correspond to the interpolation regime of the neural networks. The gradients of the fitness are calculated using tensorflow's automatic differentiation capability. The initial latent vectors are updated according to the fitness gradients and the process is repeated. After convergence, the best solution is kept as final design.

This process is able to faithfully create pixel patterns that result in distinctly different optical outputs, depending on which input mode is injected to the device. The varFDTD simulated result of the design target shown in figure \ref{fig:cwg_and_network_scheme}c is depicted in figure~\ref{fig:cwg_and_network_scheme}b, showing the successful implementation of our example target, the anti-diagonal exchange matrix, with equal phase for light injected in each of the input ports. This method is also capable of addressing the phase of each matrix element, as will be demonstrated in the following sections. Transmission matrices are depicted as colorful sets of $3 \times 3$  blocks, where the phase and amplitude of each complex matrix element are mapped respectively on the hue and saturation of an HSV color space \cite{nikkhah2024inverse}. 

\section{Further optimization of design performance}

\subsection{Inverse Design Performance Metrics}

Inverse design performance is assessed using varFDTD simulation of the optimized patterns on a test-set of 1000 Haar random unitary transmission matrix targets. A useful metric to carry out this analysis is the amplitude fidelity. The fidelity compares both the target and resimulated complex transmission matrices, with a perfect match between the two yielding a fidelity of 1. 
We calculate the amplitude fidelity using the following formula taken from literature \cite{taballione2021universal}, $\text{F} = 1/N \, \left[\text{Tr}(|U^{*}|\cdot |U_{\text{sim}}|)\right]$ where $\text{U}^{*}$ represents the conjugate transpose of the target complex transmission matrix, U$_{\text{sim}}$ is the re-simulated result, and $N$ refers to the number of modes in the given system, for our device model, $N\,=\,3$.

\begin{figure}[htb]
    \centering
    \includegraphics[width = \linewidth]{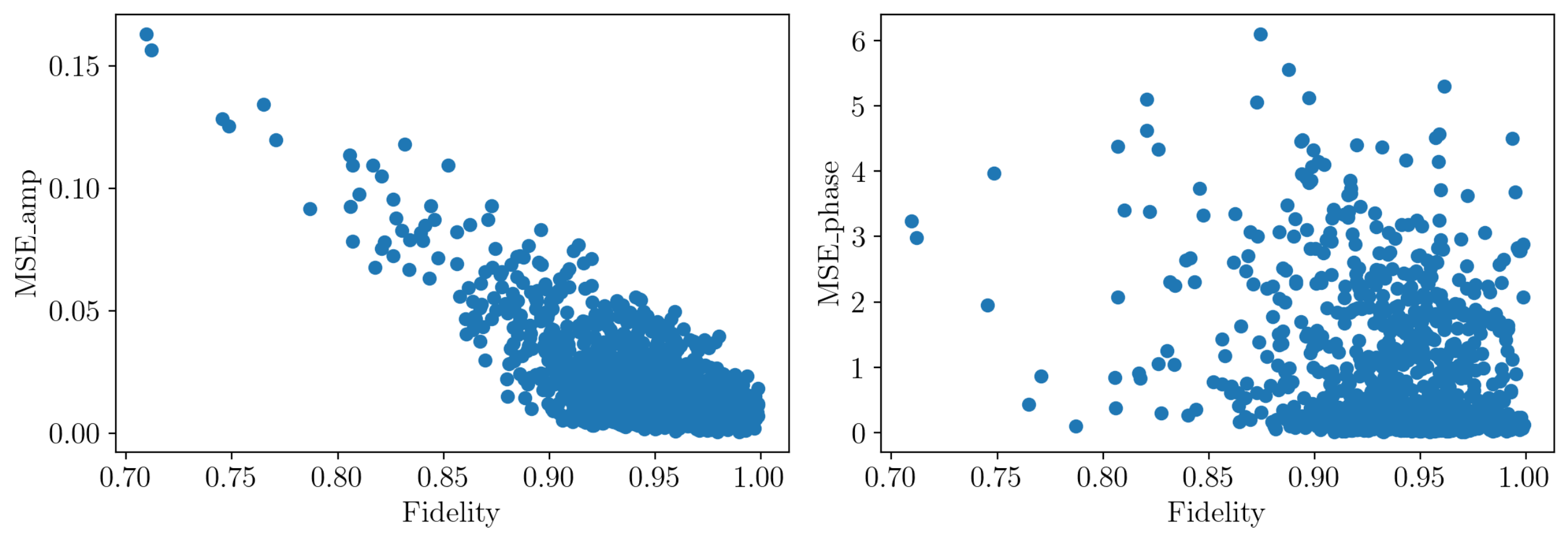}
    \caption{Correlation between amplitude fidelity and MSE for both the phase and amplitude across 1000 Haar random matrix targets. MSE is calculated between the target and resimulation matrix with the average across all 9 elements presented here. We observe a linear correlation between amplitude MSE and fidelity but a mostly uncorrelated relationship with phase MSE as we would have predicted from our fidelity formulation.}
    \label{fig:mse_v_fidelity_phase_and_amp}
\end{figure}

Amplitude fidelity as a metric is generally insensitive to the phase agreement between the two matrices under study, as demonstrated in figure\,\ref{fig:mse_v_fidelity_phase_and_amp}. While many applications do not require exact phase recreation, or are able to compensate with the addition of phase shifters at input and output ports, in some cases accurate recreation of the correct output phases is crucial. In such applications, it is instead useful to analyze the mean squared error (MSE) function between target and resimulated phases to assess network performance. The MSE values presented hereafter are calculated across all 9 matrix elements. The phase MSE was furthermore taken using the closest phase difference modulo $2\pi$ and was normalized by dividing by a factor $\pi^2$. In our analysis we consider mainly the amplitude fidelity as a measure of performance for consistency with other works in the field, but will refer to the other metrics where these are most relevant. 

\subsection{Dataset optimization}

After training both networks on the initial dataset (see above), the average amplitude fidelity achieved is around $0.86$. While this means in general a good agreement between design target and optimized solution is reached, it also means that the inverse designed patterns still deviate distinctly from the expected results. We therefore invest in optimization of the training data before assessing the performance of the inverse design approach.

\subsection*{Iterative Data Improvement} 
\begin{figure}[t!]
    \centering
    \includegraphics[width = \linewidth]{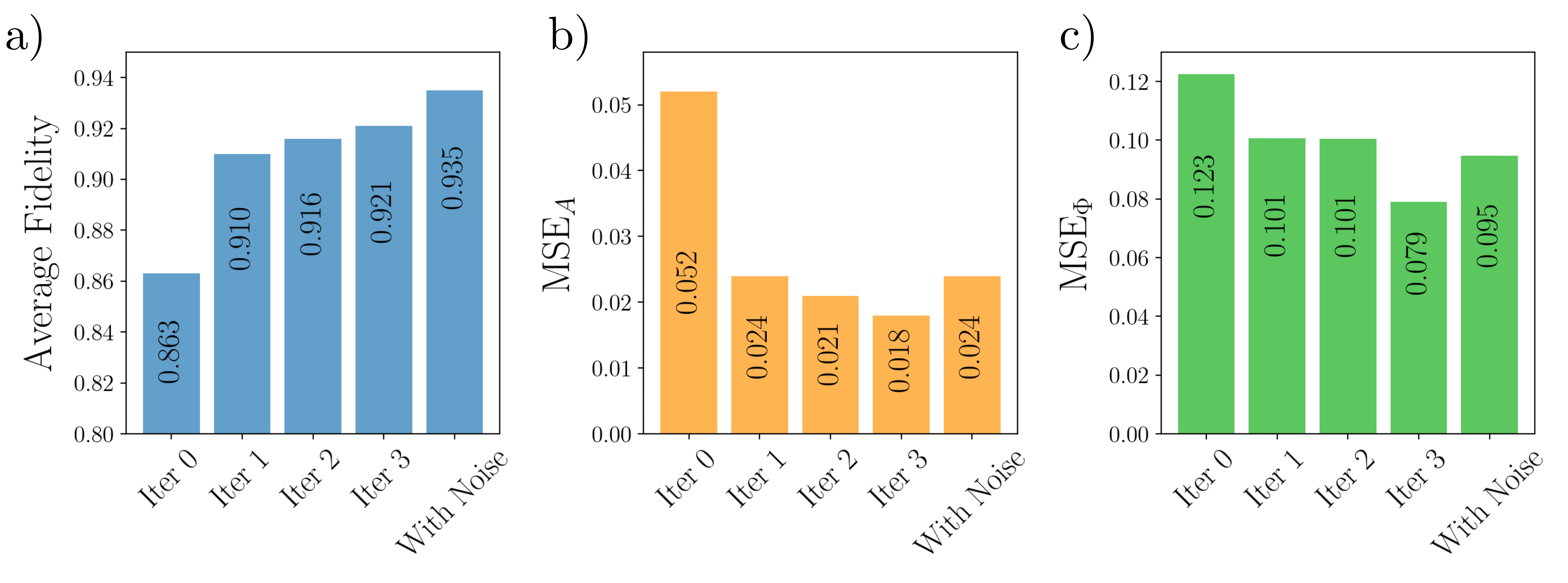}
    \caption{Average fidelity and MSE values during the process of iterative improvement. Values presented are averages for resimulation of 1000 pixel patterns produced to implement random unitary targets. Each improvement loop contained 2000 patterns which are in turn doubled by y-symmetry.} 
    \label{fig:itt_improv_error_rate}
\end{figure}

As a first measure for dataset optimization, we iteratively extend the training dataset using the results from the inverse design process itself, to improve the design fidelity. Essentially, the idea is to let the neural network predictor learn from its own mistakes \cite{blanchard-dionneSuccessiveTrainingGenerative2021, dinsdale2021deep}. 
Using the inverse design pipeline, we predict pixel patterns for 2000 random unitary targets. The transmission matrices of the designed patterns are simulated by varFDTD, and these results are appended to the initial training dataset. Subsequently, the models (both, forward predictor and WGAN-GP) are re-trained using the now extended dataset. 
We repeat this process three times, and benchmark the inverse design quality using 1000 Haar random unitary targets after each iteration. After three iterations, we observe a notable improvement in inverse design performance, with an average amplitude fidelity now of $0.92$, as shown in figure~\ref{fig:itt_improv_error_rate}.
This performance is comparable to fidelities obtained by similarly available reconfigurable technologies \cite{taballione2021universal,de2022high, taballione202320}.

\subsection*{Expansion with Random Pixel Patterns} \label{sec:random} The CWGs exhibit a well defined small number of modes and randomly generated patterns result in high device throughput, as opposed to for example MMIs where the number of modes inside the device is much higher and therefore high-throughput solutions are more sparse \cite{dinsdale2021deep}. This property in principle allows us to rapidly add patterns to the training dataset without the need to optimize for device throughput. It may therefore seem appealing to simply append as many random patterns as possible to the training set.  However, in our studies we have found a more methodical approach is necessary, as is shown in more detail in supplementary information figure~\ref{fig:SI_supp_data_percentage_network_losses} and simply adding more random data does not improve the network performance.

\begin{figure}[htb]
    \centering
    \includegraphics[width = 0.8\linewidth]{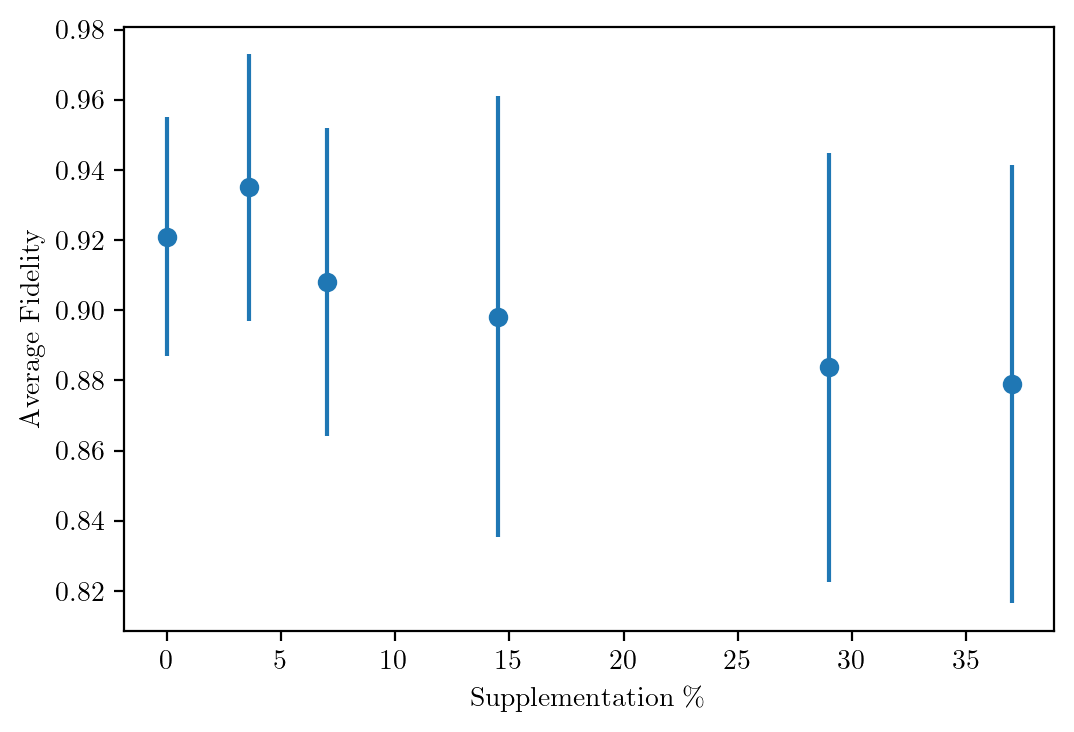}
    \caption{After dataset augmentation with varying percentages of random noisy data, each network was presented with 1000 random unitary targets for which the average fidelity is plotted above.}
    \label{fig:SI_supp_data_percentage}
\end{figure}

The reason that purely random datasets are found to be ineffective is attributed to the collective action of the pixel patterns. Within the large space of possible solutions, there are many that are not particularly effective in implementing a transmission matrix, as pixels do not sufficiently act in concert or even counteract each other. By including randomly generated data we modify the statistical distribution, which starts containing more and more of these ''counterproductive'' patterns. Part of the forward model's capacity is then used to fit these useless noise cases which are not beneficial for the later inverse design task. The WGAN latent space on the other hand will become less efficient in restricting the optimization to useful geometries, since it is also trained on the same data now containing totally random patterns, which are therefore part of its latent space as well. Consequently it is foreseeable that the addition of too many random patterns will result in a decrease of overall inverse design accuracy, which is demonstrated in the supporting information figure\,\ref{fig:SI_supp_data_percentage_network_losses} and \ref{fig:SI_supp_data_percentage}. 

Adding a small amount of totally random pixel patterns slightly improves the inverse design capacity, yielding decreased validation losses, as well as an increase in the average fidelity from $0.921$ to $0.935$, as further shown in figure~\ref{fig:itt_improv_error_rate} (cases ``with noise''). This trend is consistent with other work where a small amount of randomness was found to improve performance \cite{klokkou2023deep}. We attribute this improvement to a better prediction of edge case geometries. As pattern optimization happens inside the WGAN latent space, the inverse design is restricted to the statistical distribution of the initial training data. Since the initial data has a bias towards low pixel numbers it may actually be beneficial to also add totally random pixel patterns in order to diversify the training set and make the forward network better deal with patterns at the edge of its latent space


\subsection{Pattern characteristics}

\begin{figure}[ht!b]
    \centering
    \includegraphics[width = 0.6\linewidth]{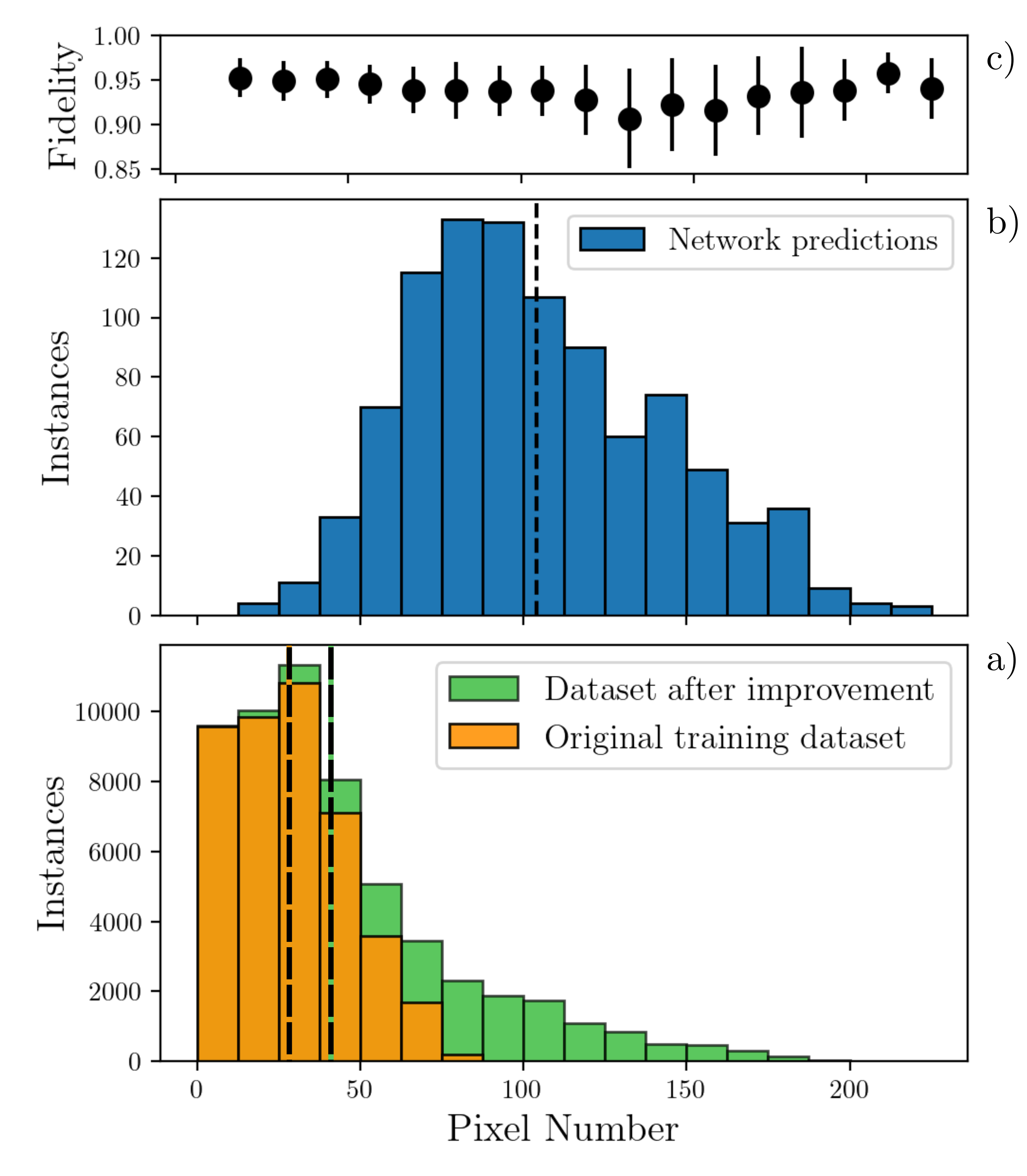} 
    \caption{(a) Distribution of pixel numbers contained in patterns from the initial 'brute force' training dataset (orange) and final training dataset after iterative improvement and 2.5\% random patterns (green). (b) Pixel histogram of predicted patterns for 1000 random unitary targets. (c) Corresponding average fidelities for each bin of the predicted patterns, with error bars representing the standard deviation from the mean. Dashed lines: average values of 28 pixels for initial training data and 41 pixels for final training data.}
    \label{fig:total_training_data_pixel_number}
\end{figure}

The bottom panel in Figure\,\ref{fig:total_training_data_pixel_number} shows histograms of the distribution of active pixels in the initial datasets used for training (orange) and the augmented datasets after three iteration loops (green). The pixel histogram for patterns predicted by the network for a validation dataset of 1000  target transmission matrices are shown in the middle panel (blue). The corresponding average and standard deviation of the fidelity for the predicted patterns is presented in the top panel of the figure. As the training data is generated in an iterative manner and intermediate steps are stored, there is an strong emphasis in the training dataset for patterns contain a low number of pixels. Across the entire database used for training there is an average number of 41 pixels/pattern. 

In comparison, the average number of pixels in predicted patterns is 104 and patterns with up to 225 active pixels are found. Implementing the target transmission matrices requires predominantly geometries containing more pixels than the average number from the training data. We can observe that as the number of pixels increases above 50, the standard deviation of the fidelity increases quite significantly. An increase in network error rate will lead to a greater spread of results and thus a larger standard deviation. This trend is attributed to the forward network performing less accurately on geometries that are underrepresented in its training data, which could explain some of the errors in the inverse design. More discussion on the infuence of forward network accuracy on fidelity will be given below in Section~\ref{sec:unitary}.


\begin{figure}[t!]
  \centering
  \includegraphics[width = \linewidth]{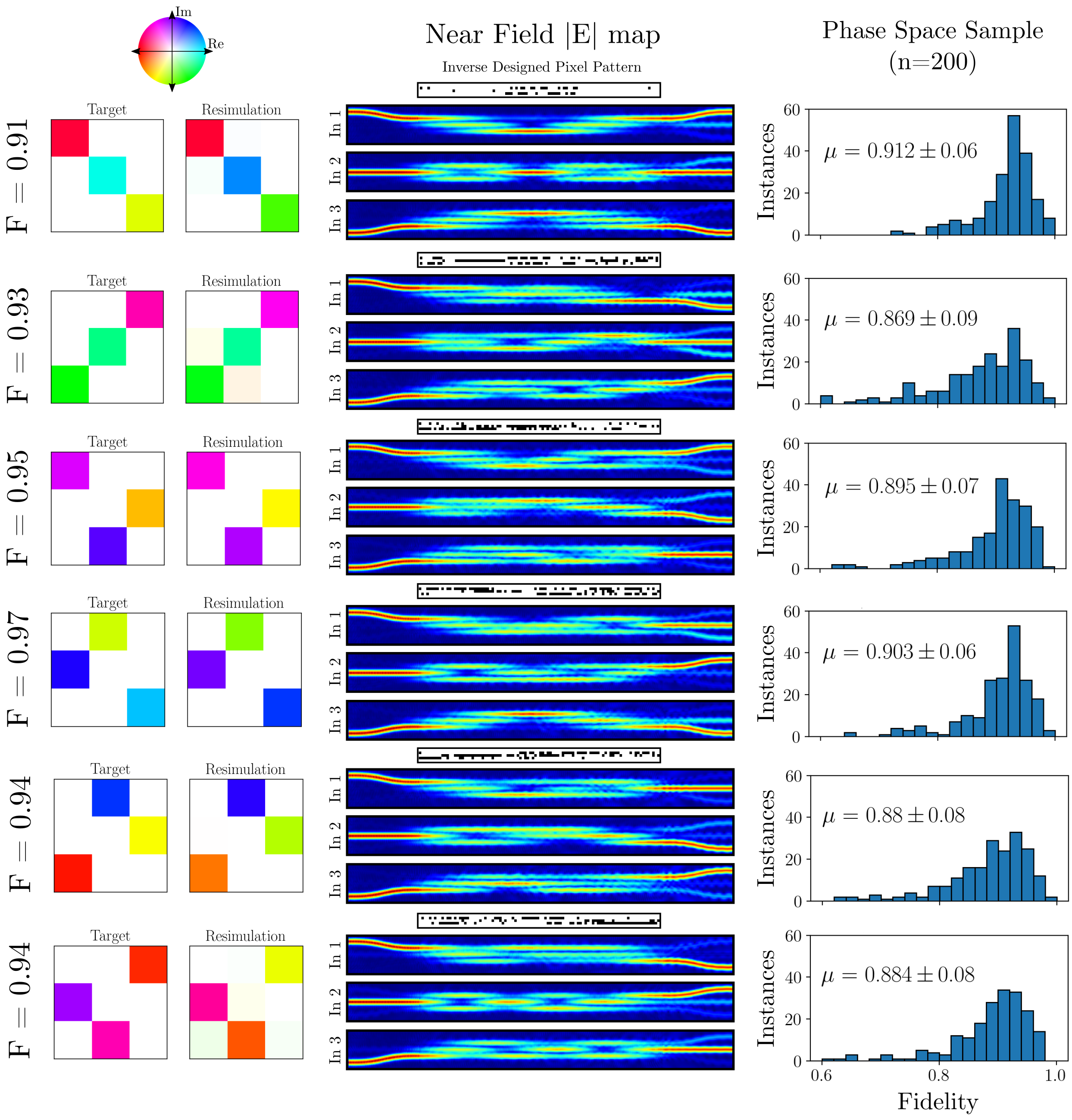}
  \caption{Performance of predicted patterns for the implementation of all available permutation matrices of the n=3 waveguide array. 200 random phase shifts are applied to each matrix element, allowing for in depth sampling of the available phase space to determine a maximum fidelity for phase insensitive applications. Average fidelities of around 0.9 are achieved for all targets, with peak fidelity values $>0.97$ for all targets.} 
  \label{fig:3x3_results_grouped}
\end{figure}


\section{Results on Transmission Matrix Design}

After the iterative dataset expansion, the addition of random patterns, and the training of the network on this final dataset, we proceed to carry out a benchmark of our approach for the inverse design of pixel patterns for a variety of coupled waveguide transmission matrices. These can be broadly split into two groups: permutation matrices, which are a class of orthogonal matrices where individual input and output ports are connected without port mixing, and unitary matrices, providing the most general input-output relationship. In the following we discuss these cases in more detail.

\subsection{Permutation Matrices}
The first class of matrices under study contains those which guide light in a one-to-one fashion, whereby light injection into each input waveguide results in transmission through only one unique output. For a transmission matrix with $n\hspace{-0.04cm}\times\hspace{-0.04cm}n$ elements, there will be $n!$ unique permutation matrices, each of which for our device can be seen in figure\,\ref{fig:3x3_results_grouped}. 
An example phase shifted matrix is presented for each permutation target as well as its associated varFDTD simulation of the near field electric field intensity.
These matrices are a useful test to ensure that the network has generalized to a point where it is able to predict patterns for targets which lie on the very edge of the training geometry space (those confining the output electric field fully to only one port), and to check the ability of inverse designed patterns to produce distinctly different optical outputs for each input mode. In our design challenge, we chose to set the phase of each non-zero matrix element as a free parameter which can be chosen to obtain the best solution in amplitude. This approach appears useful since i) in many use cases, the phase response of an optical switch is not critical and ii) if necessary, an optical phase shifter at each of the outputs is sufficient to rebalance the output phases in the device.

Results shown in figure\,\ref{fig:3x3_results_grouped} demonstrate accurate reproduction of both the amplitude and phase distribution for a selection of matrices, demonstrating that a well performing solution can be achieved for all of the orthogonal permutation matrices in the $3 \times 3$ group. Here solutions were selected on the basis of combining fidelity with a low phase MSE. Pixel patterns as well as local field distributions for each of the input ports are presented for completeness for each of the devices, providing some insight in the internal structure of these solutions. For example, for the identity matrix at the top, we see that the pattern of perturbations is predominantly placed in the middle of the device along its length. This pattern transforms the original device, which had no self-imaging capability, to produce a highly symmetric self-imaging of each of the three inputs onto the output plane. 

Anti-diagonal and more complex permutations are seen in the other five diagrams, where the common denominator is that the field profiles typically show only a few bounces between input and output, indicative of the coupling length of the CWG of around 10~$\mu$m and the fact that the modal basis is very small. Weak perturbations therefore tend to couple modes in an adiabatic way, without giving rise to strong scattering events or reflections \cite{Bruck2016}.  


A sample of fidelities across the available phase space for each permutation matrix is presented in the right column in figure\,\ref{fig:3x3_results_grouped}, in which 200 random phase shifts are applied to each element in the target matrix. Average fidelities are achieved of around 90\% across the phase space sample for all 6 target matrices. When output phases can be freely adjusted to maximize the fidelity, the phase space sampling technique permits significant improvements to the fidelity, resulting in fidelities in excess of 0.97 for all targets.


\begin{figure}[h!]
    \centering
    \includegraphics[width = \linewidth]{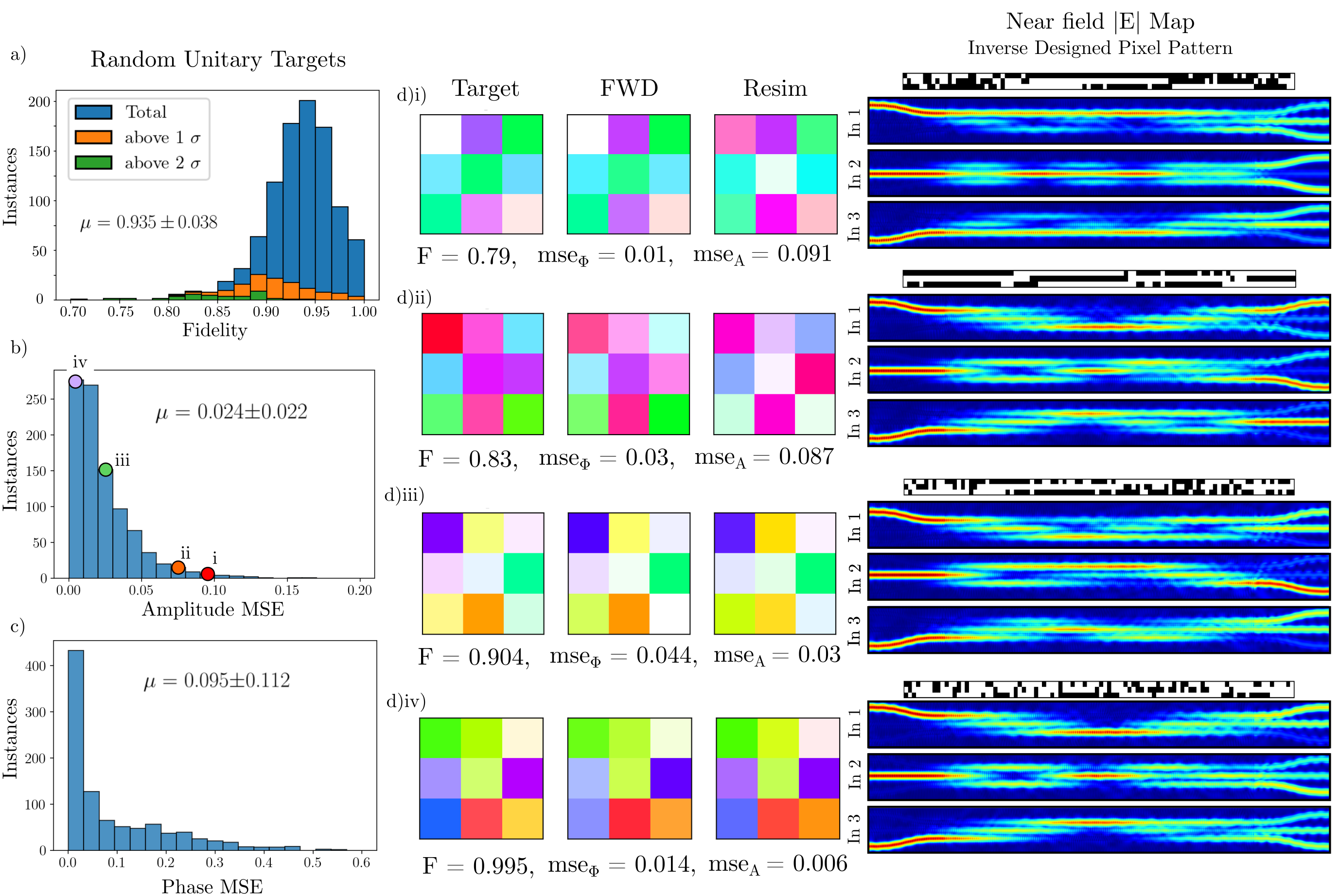}
    \caption{a) Distribution of fidelities for the reproduction of 1000 random unitary targets. Highlighted are instances which fall outside 1 and 2$\sigma$ of the average MSE between forward network prediction and resimulation, indicating a failed forward network prediction. b) and c) show the distribution of MSE values between target and resimulation for these same targets. d) 4 example unitary targets sampled from across the achieved fidelity distribution. We present the target, forward network prediction of the inverse designed pixel pattern, and a resimulated transmission matrix. To the right hand side we plot the respective near field electric field intensity map for a device programmed with the predicted pattern for each target matrix.}
    \label{fig:network_results_random_target}
\end{figure}
  

\subsection{General Unitary Transmission Matrices}\label{sec:unitary}
Next to the design of orthogonal permutation matrix targets, which pose a useful design task to ensure that we have precise control over every individual matrix element, in many real world use cases such as optical computing and quantum information processing, mixtures of inputs are required. Unitary matrices represent a particularly useful class of targets representing the most general operations available in the multi-port system. These targets must satisfy the condition $\text{U}\text{U}^{*} = \text{I}$, where U$^{*}$ represents the conjugate transpose of the target matrix and $I$ is the identity matrix, corresponding to a lossless device used for target matrices. 

Figure\,\ref{fig:network_results_random_target}a) shows the fidelity distribution for network predicted patterns from a set of 1000 random unitary targets drawn from the Haar distribution. We find that the neural network is able to reach design targets with good fidelity of $0.935 \pm 0.038$. This is reflected in the convergence to low values below 0.05 in the amplitude MSE. The distribution of MSE values for phase shows that the prediction of correct phase distributions is more challenging, which will be discussed in Section~\ref{sec:discussion} in more detail. A sub-selection of four matrix targets from across the fidelity distribution are presented in figure \,\ref{fig:network_results_random_target}b) labelled as i) - iv), as indicated by the colored dots in figure \,\ref{fig:network_results_random_target}b), showing the target matrix, the forward network prediction and varFDTD resimulation result. Pixel patterns and near field maps are present the corresponding microscopic configurations of the four patterned devices under study. It can be seen that for all four examples, the forward network prediction agrees well with the target matrix. Main differences can be seen between the network prediction and the varFDTD result, suggesting that an important factor in the fidelity may be the accuracy of the forward network. 

To verify the influence of the forward network error, we calculate the MSE between forward network prediction and varFDTD calculation. The fidelity distributions are obtained for parts of the dataset where the MSE lies above one and two standard deviations (labelled as $1\sigma$ and $2\sigma$), given by respectively the orange and green histograms in figure~\ref{fig:network_results_random_target}a). This analysis confirms what is seen in the examples, namely that the tail of lower fidelities is correlated with a poor accuracy of the forward network prediction.

The examples furthermore indicate that phase and amplitude errors are not related. Results with low phase error can be found for a low fidelity and \textit{vice versa}.
Figure\,\ref{fig:network_results_random_target}d)iii) for example shows visually a good match to the target unitary colors due to low phase error, however the final amplitude fidelity remains below average at 0.9 owing to a disagreement between amplitude values. The near field maps and pixel patterns of random unitaries are not easily understood through intuition, while an underlying phenomenology would be of interest this goes beyond the scope of our study. We do observe the emergence of larger section of connected pixels forming lines, which may be a strategy of the network to achieve large phase shifts in certain matrix elements. Some more discussion on the phase structure of the CWGs is presented in Section~\ref{sec:discussion}.



\subsection{Hadamard and Fourier Matrix}
Complex Hadamard matrices play an important role in quantum information theory. They have been used to tackle a number of problems including the development of spin models \cite{nomura1994spin} and analogue quantum simulators \cite{leung2002simulation}. They have also helped establish mathematical frameworks to construct bases of unitary operators and maximally entangled states. Fourier matrices can be used to apply a discrete Fourier transform to a signal through matrix multiplication, making them of particular interest in optical signal processing and computation. For a 3$\times$3 matrix, all complex Hadamard matrices are equivalent to the Fourier matrix, $F_{3}$ \cite{tadej2006concise}:

$$ F_{3} = \begin{bmatrix}
1 & 1 & 1 \\
1 & \omega & \omega^{2} \\
1 & \omega^{2} & \omega 
\end{bmatrix}  $$

\begin{figure}[t]
    \centering
    \includegraphics[width = \linewidth]{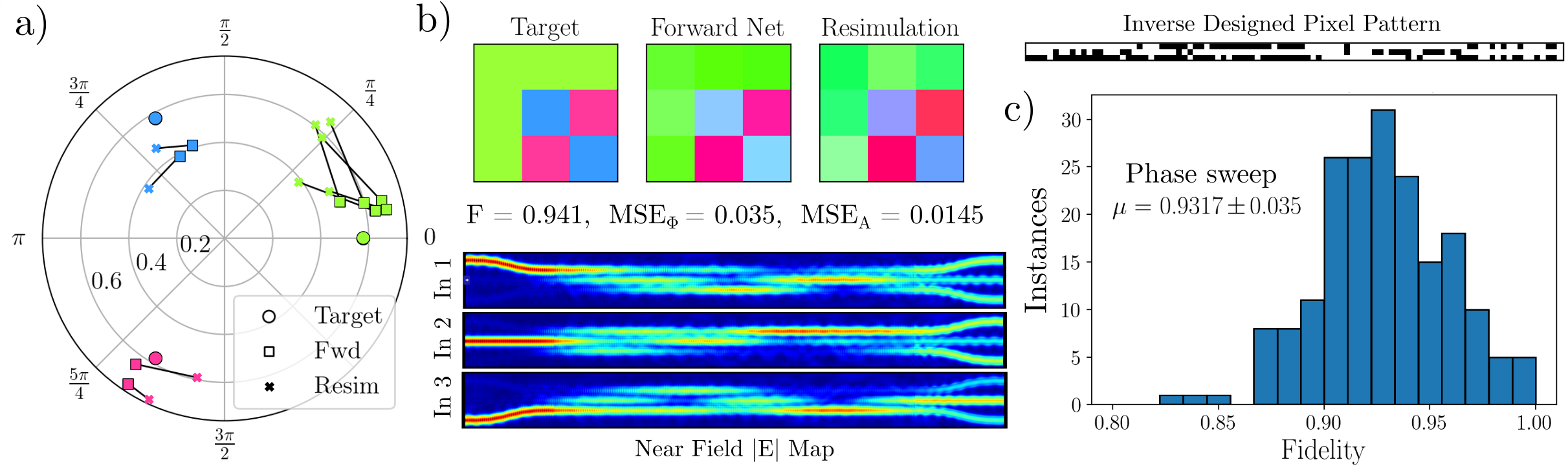}
    \caption{a) Polar plot comparing each matrix element in the reproduction of the 3$\times$3 Hadamard and Fourier matrix (shown in sub figure b). Datapoints are color coded to their associated target value, and black lines indicate the distance on the complex plane between forward network predictions and their resimulated matrix values. c) Shows again a phase sweep, in which a global phase shift is applied to the target matrix, retaining the same intensity distribution resulting in a maximum achievable fidelity in excess of 0.99.}
    \label{fig:Hadamard_recreation}
\end{figure}

Where $\omega = \text{exp}\left(2\pi i/3\right)$. It is of interest to check the existence of a solution for this specific matrix operator. As $F_3$ is a unique matrix, we again give the network some more flexibility in the boundary conditions by allowing a single global phase factor within all matrix elements. A global phase factor can easily be factorized out and compensated in an optical system. It makes sense to include this degree of freedom in the matrix to find the best working point of the CWG under study, taking into account both amplitude and phase MSE. Figure\,\ref{fig:Hadamard_recreation}a) and b) shows the forward network prediction and resimulated result when attempting to implement the $F_{3}$ matrix. An amplitude fidelity of 0.94 is achieved for the selected solution, with a phase MSE of 0.035 and amplitude MSE of 0.0145. The polar plot in figure\,\ref{fig:Hadamard_recreation}a) is used as a graphical depiction of the accuracy for each individual matrix element, data points are colour coded corresponding with the matrix element they represent, with black lines between associated network predictions and resimulated points representing their separation within the complex plane. 

The full histogram of fidelities for a range of global phase values between $-\pi$ and $\pi$ is shown in figure\,\ref{fig:Hadamard_recreation}c) and allows to identify an average fidelity across this phase sample of 0.93, while we again are able to retrieve a peak fidelity of $>0.99$ but at a higher phase MSE. 

\section{Discussion}\label{sec:discussion}
Our work shown here presents coupled waveguide arrays as an alternative to other commonly used reconfigurable platforms such as interferometer meshes or multi-mode based devices for programming arbitrary unitary operators. Approaches for using such devices are well established, with practical realizations demonstrated across a number of fields as discussed in the preamble to this work. Interferometer meshes in particular provide a high degree of control, allowing for the introduction of individual phase delays at single unit cells across the mesh for highly accurate matrix reproduction. There exist several mathematical models \cite{reck1994experimental, clements2016optimal} which allow for the calculation of where these phase shifts must be applied for arbitrary matrix decomposition. Despite promising results, small insertion losses at each interferometer will sum together, leading to non-negligible losses across the whole device. Furthermore, when targeting large matrices, the total device footprint becomes large when compared to alternatives. Adding to this, many approaches use thermo-optic modulators to introduce the required phase shifts, which add further to this size,  introduce electronic overhead and regulating circuits, and produce thermal instability across the chip.

Multi-mode structures such as MMIs and devices which are built around multi-mode waveguides are the most compact technology discussed here. Extra degrees of freedom introduced by the inclusion of different optical modes allow for highly compact matrix decomposition, however, this is not without its drawbacks. Coupling out of the device becomes challenging as any modal-mismatch can introduce significant losses, often requiring some signal pre-processing through techniques such as beam shaping. The need for accurate coupling between modes also now imposes a requirement for high degrees of spatial accuracy in the programming of such a device. Coupled waveguide arrays strike a midpoint in device footprint, necessitating only slightly more space in the direction of light propagation to allow a round trip between outermost waveguides if aiming for the smallest device possible. In the plane perpendicular to the light injection axis their size remains comparable, if not smaller than MMI type devices. This compact form-factor is significantly smaller than comparable interferometer meshes, and remains within a consistent modal basis, introducing no losses from coupling between modes at the output waveguides making it an attractive option for space-limited applications.


While in general the network performance in our work is strong, there are some examples where the resimulated results are not a good match to the target matrix. By comparing the forward network predictions to resimulated results we may check if forward network predictions remain valid. Using a simple MSE discriminator, instances which fall outside one standard deviation from the mean can be identified and rejected, highlighting  cases where the forward network has failed to accurately predict the transmission matrix for a given pattern. The fidelity distribution with cases above 1 and 2$\sigma$ highlighted were shown in figure\,\ref{fig:network_results_random_target}. For these designs, the network appears to have strayed from the interpolation regime of the forward network, resulting in large errors in the predicted transmission matrix. Rather unsurprisingly, almost every fidelity below 0.85 is a result of a failed network prediction. Excluding these from our statistical analysis the average fidelity now raises to $0.94\pm0.029$. In the majority of cases however, it appears that the optimization of the geometry in the WGAN's latent space and the addition of the latent constraint successfully limits designs to the surrogate network's validity region. Iterative data improvement may also be continued beyond the three cycles in this study, which apart from making the network learn from its own mistakes allows for augmenting the training data toward larger pixel numbers. The performance gain has to be traded off against the effort needed as the additional improvement for each cycle is expected to saturate.

\begin{figure}[htb]
    \centering
    \includegraphics[width = \linewidth]{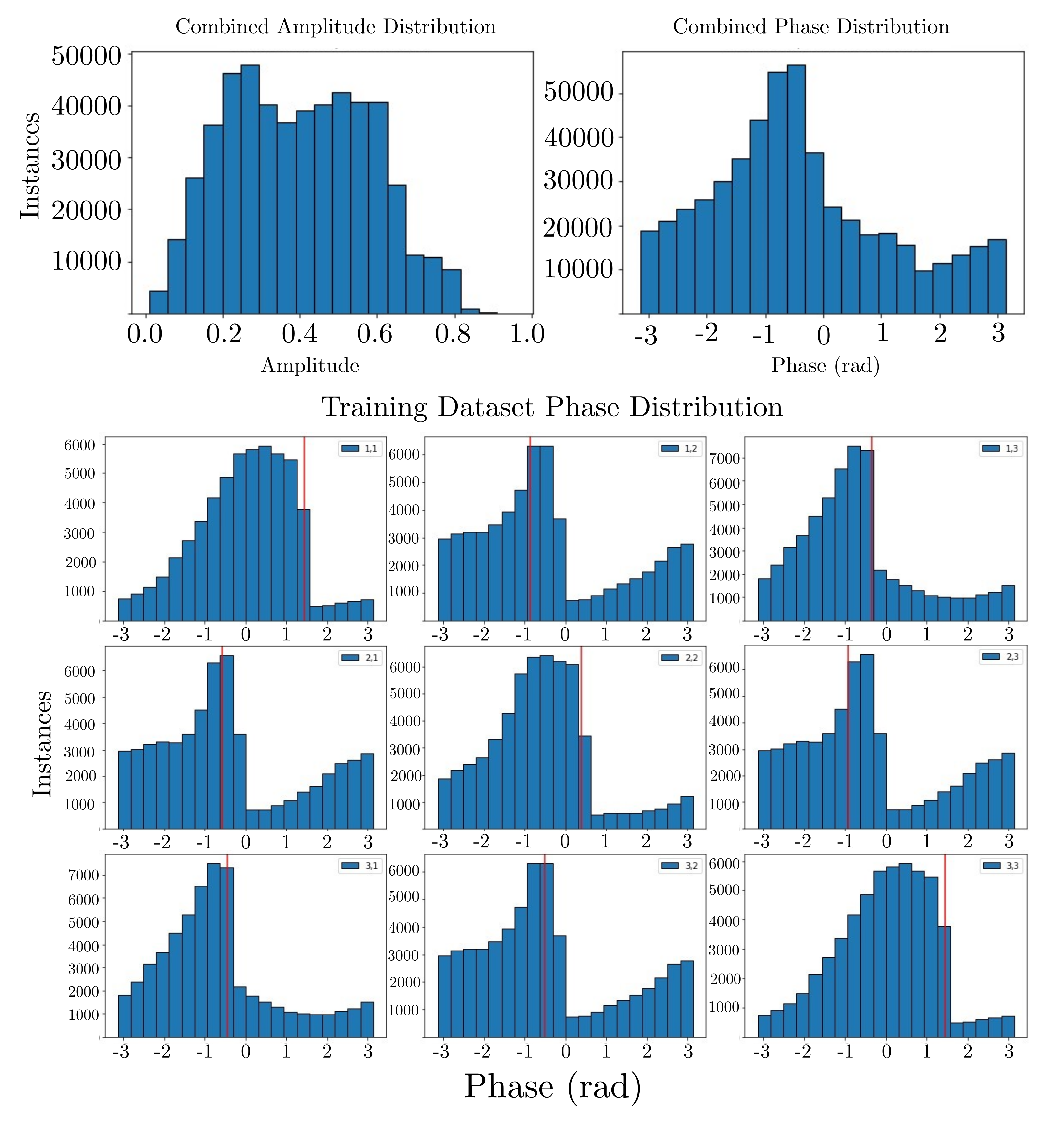}
    \caption{Phase and intensity distributions across each matrix element in the training dataset. Amplitude values show a mostly normal distribution centered around 0.33 as we would expect for a n=3 waveguide array. Phase values are not addressed in data generation and therefore tend to smaller phase shifts as these will introduce fewer losses. Below we see the phase distribution per matrix element further demonstrating this behaviour. The red line indicated the phase of an unperturbed device, where it should be noted phase is wrapped at $2\pi$ so the ride in instances to the right hand side is in fact the end of the tail to the left.}
    \label{phase_and_amp_training_data_distribution}
\end{figure}

The network is trained on the full complex field, therefore one might expect similar performance for phase and amplitude, study of our respective average MSE values shows this is however, not the case. One hypothesis for the cause of this discrepancy originates in how brute force optimization targets are generated. During the data generation stage, the phase of light at the outputs is unconstrained and given no optimization weighting, we asses the success of each pixel iteration solely on the intensity at the output ports. Intensity optimization targets are randomly selected from a uniform distribution, resulting in the retrieval of a normal distribution of possible amplitude values centered around 0.33 as we expect for a 3-waveguide system. Conversely, analysis of the phase distribution for the same data (shown in figure\,\ref{phase_and_amp_training_data_distribution}) shows distinctly different behaviour. Unsurprisingly we record a peak for phase values matching to those of an unperturbed device, however, because the optimization is not pushed to suggest pixel patterns with ''extreme'' phase delays approaching $2\pi$, the dataset becomes highly biased towards smaller phase shifts. Although these phases can be accessed by switching extra pixels, there is no incentive for the optimization in doing so, as excess pixels come at the cost of increased scattering losses and lower transmissions, making these patterns more likely to be rejected. Because of this, the final database of pixel patterns represents the available intensity space well, but may fall short when describing the phase space relating to longer phase delays. Consequently it follows that the networks ability to accurately predict patterns to implement such phase delays will be inhibited, explaining the differing MSE results for phase and amplitude we observe. 

\begin{figure}[htb]
    \centering
    \includegraphics[width = 0.7\linewidth]{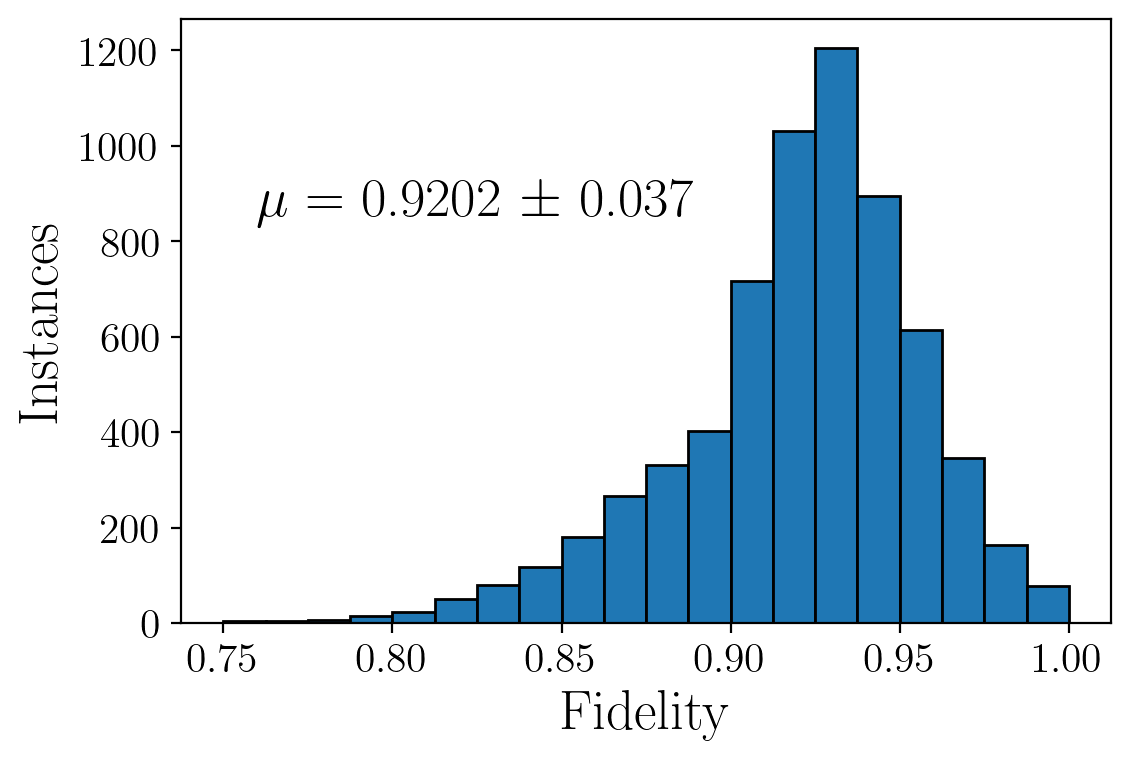}
    \caption{Amplitude fidelity distribution taken from forward network predictions for the validation dataset of pixel patterns. This prediction error is intrinsic to all presented results, and will become a limiting factor as the network is improved, outlining an upper limit on total inverse design accuracy.}
    \label{prediction_error_fidelity_distribution}
\end{figure}

The major limiting factor in the final generation of valid pixel patterns for user-defined matrix targets is therefore the accuracy with which the forward network can make predictions. Assessment of this will give some insight as to the expected upper limit on network performance when presented with an external design target. Figure\,\ref{prediction_error_fidelity_distribution} shows the distribution of fidelities from forward network predictions on the validation dataset discussed in section 2.3. This represents a ''best case'' scenario, as these patterns are generated in precisely the same way as the training data. Across the 6564 validation patterns we record an average fidelity of $0.92\pm0.04$, indicating that for our current predictor model, the achieved average amplitude fidelity of 0.94 constitutes what is likely a near maximal achievable value, which is limited by the network performance rather than the physical CWG system. More conventional topology optimizations starting from the endpoint of the neural-adjoint inverse design may be of interest to further converge solutions, which could be a topic of future work.


\section{Conclusions}

In conclusion, our study addresses the potential of low-loss optical phase change materials combined with coupled waveguide arrays as a promising and exciting avenue for the production of next-generation reconfigurable technologies across fields such as quantum simulation, photonic computing and optical data processing. In comparison with other approaches such as integrated circuits based around interferometer meshes or MMIs, coupled waveguide arrays can offer ultra-low-loss reversible modulation requiring no active regulation within a highly compact device footprint. The development of a robust inverse design pipeline using neural network surrogate models allows for rapid, near-real-time prediction of the complex pixel patterns, required to implement a wide range of transmission matrices within a single device model. In our work, the introduction of a Wasserstein generative adversarial network provides a crucial constraint on the gradient based optimization, limiting predicted geometries to those within the interpolation region of the forward surrogate model. Network performance is enhanced by augmenting and expanding the training dataset, initially through a process of iterative improvement, and in a final step, with the introduction of a small percentage of noisy random data. Although training of the networks is time consuming, taking a few hours with standard consumer level computer hardware, the entire inverse design process remains significantly faster than alternatives such as topology optimizations, allowing predictions on a millisecond timescale once trained.



Presented results demonstrate a high level of control over both intensity and phase of individual matrix elements. While some extreme phase relations may suggest the need to expand the waveguide geometries to longer device, therefore allowing multiple vertical passes of the light, generally performance is strong. Average fidelities of $0.935\,\pm\,0.04$ are reported, demonstrating comparable performance to other state of the art, commercially available reconfigurable photonic technologies and further reinforcing the validity of this approach. The fidelity obtained in our study is currently limited by the neural network performance rather than the coupled waveguide system itself and further improvement may be possible in future work. The neural-adjoint design platform introduced is highly versatile and requires minimal overheads to allow functionality across different devices and geometries, showing promise for integration with a range of future optical technologies.

\bibliographystyle{ieeetr}
\bibliography{bibliography}

\newpage

\section*{Supplementary Information}
\beginsupplement


\begin{figure}[htb!]
    \centering
    \includegraphics[width = 0.95\linewidth]{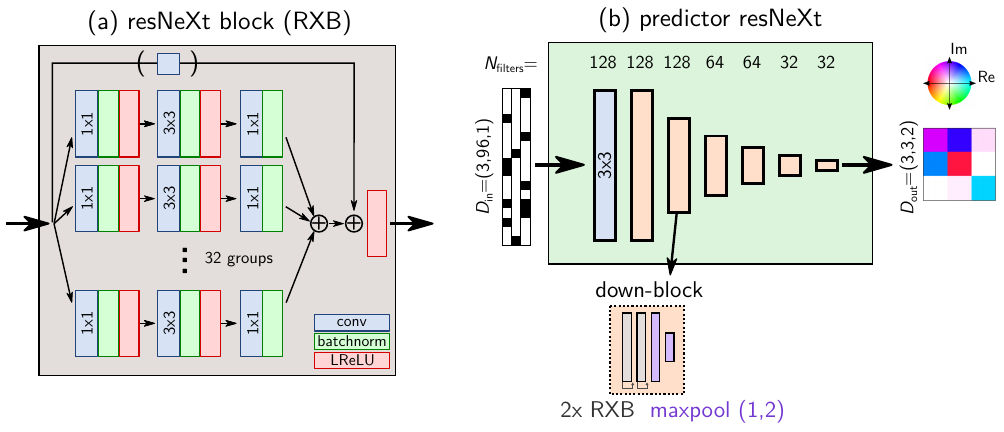}
    \caption{
    Detailed sketch of the forward network architecture. The transmission predictor is composed of ResNeXt blocks, using grouped convolutions (a). We use a cardinality of 32 and a bottleneck width of 4. For similar computational cost and network complexity, they are known to be more efficient than classic residual blocks \cite{xieAggregatedResidualTransformations2017}. The forward model (b) is trained for 100 epochs using a simple plateau learning rate reduction with a factor of $0.5$. Every 25 epochs we furthermore increase the batchsize by a factor of 2, from a starting value of 16, following the suggestion of \cite{smithDonDecayLearning2018}.
    }
    \label{fig:forward_NN_sketch}
\end{figure}

\begin{figure}[h!]
    \centering
    \includegraphics[width = 0.95\linewidth]{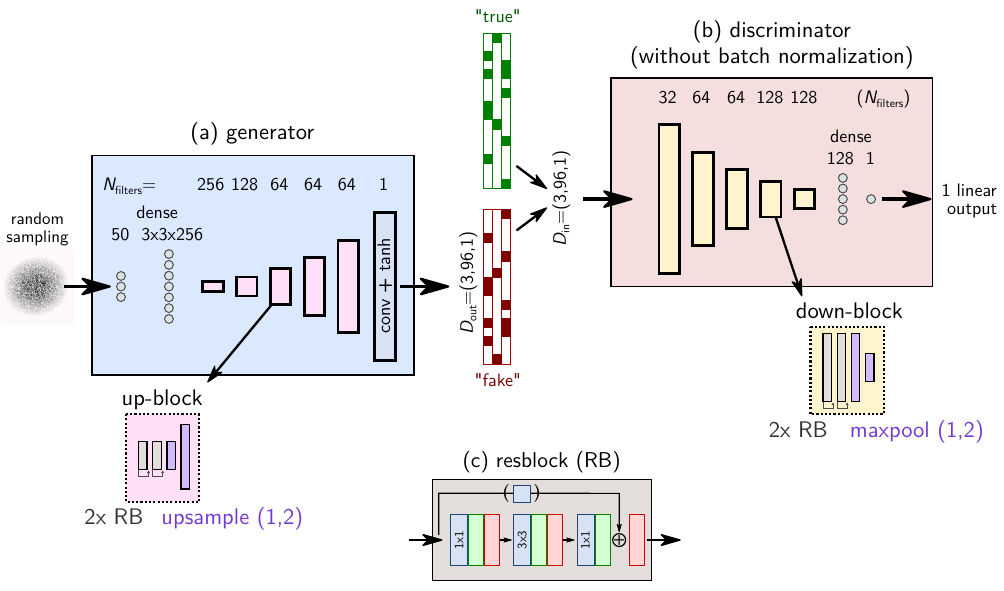}
    \caption{
    Detailed sketch of the WGAN-GP generator (a) and discriminator (b) architecture, following closely the layout of the original implementation \cite{gulrajani2017improved}. In particular, we use conventional resnet blocks (c) and no batch normalization in the discriminator. The latter is trained for 3 steps for a single training step of the generator. We train during 15 epochs first with a batch sizes of 64, followed by 15 additional epochs with a batchsize of 128. Both networks are trained using the Adam optimizer \cite{kingmaAdamMethodStochastic2014} with a fixed learning rate of $0.00005$.
    }
    \label{fig:WGAN_sketch}
\end{figure}

\begin{figure}[h!]
    \centering
    \includegraphics[width = 0.7\linewidth]{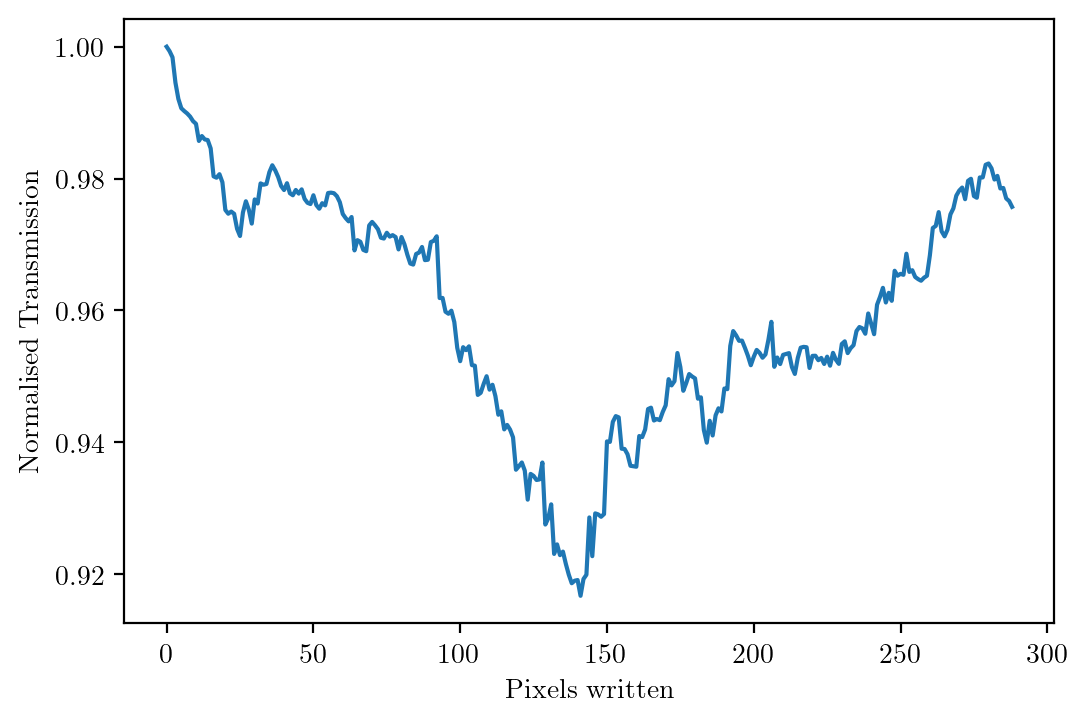}
    \caption{Total device transmission normalised to the unperturbed throughput as a function of number of pixels switched. Switching pixels introduces reflections and scattering losses at the pixel boundaries, decreasing transmission. As $>50\%$ of pixels are switched the number of these boundaries decreases and transmission increases. The maximum recorded drop in transmission is 8\%.}
    \label{scattering_loss_v_pixel_number}
\end{figure}

\subsection*{Brute force iterative optimization}

A brute force iterative optimization is used to generate training data for training the network presented above. The algorithm is a simple approach which can be applied to any multiport device. To begin the device structure in its unperturbed state is set up using a commonly available varFDTD software. A random splitting ratio and input waveguide is selected before simulating the device to determine the initial output characteristics, a weight is then calculated using a common mathematical approach, for presented training data the weight function is as follows:

\begin{equation}
    W = \sum_{i = 1}^{n}1-\Delta \text{T}_{i}
\end{equation}

Where $n$ is the number of output ports, $\Delta$T represents the difference in transmission between the target and current geometry. As the output approaches the target the weight function is maximized and therefore at each iteration we can compare the performance of the pattern on average. This approach is sufficient for a simple single port optimization, however, if a multi-port optimization is preferable then a further summation over the input waveguides would be necessary also. 

After the initial weight has been calculated a random pixel is selected and its state switched between crystalline or amorphous. A new weight is now calculated, if the weight has increased we know the transmission has moved closer to the target and the perturbation can be kept, if not the perturbation is reversed. Each time a pixel is saved a sweep across all input waveguides is carried out, allowing for the construction of the full device transmission matrix. Initial testing showed that choosing pixels at random to perturb, decreased simulation time and increased the optimization accuracy for a set number of iterations.

\begin{figure}[t]
    \centering
    \includegraphics[width = 0.7\linewidth]{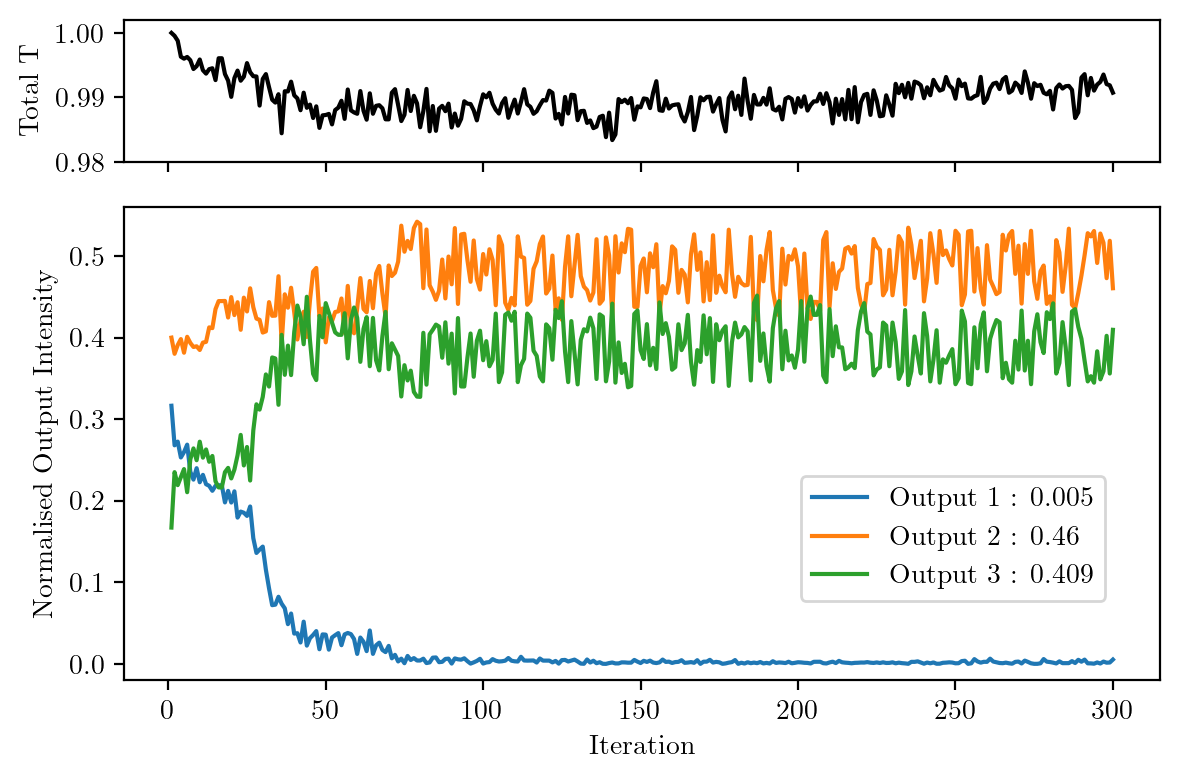}
    \caption{Example brute force optimization of a single input for a 3$\times$3 waveguide array. The target splitting ratio (0\%/50\%/40\%) is approached within 100 iterations before becoming stuck in a local minima. Total device throughput, shown in the top panel remains high as pixels are written, dropping only by 0.6\% across the entire optimization.} 
    \label{fig:SI_Brute_force_optimisation}
\end{figure}

Figure\,\ref{fig:SI_Brute_force_optimisation} shows an example optimization for a 0\%/50\%/40\% splitting ratio for light injected into input channel 1. This demonstrates both the negligible scattering losses as more and more pixels are written onto the device using our platform, as well as the tendency of such an approach to become stuck in a local minima.

\begin{figure}[htb]
    \centering
    \includegraphics[width = 0.7\linewidth]{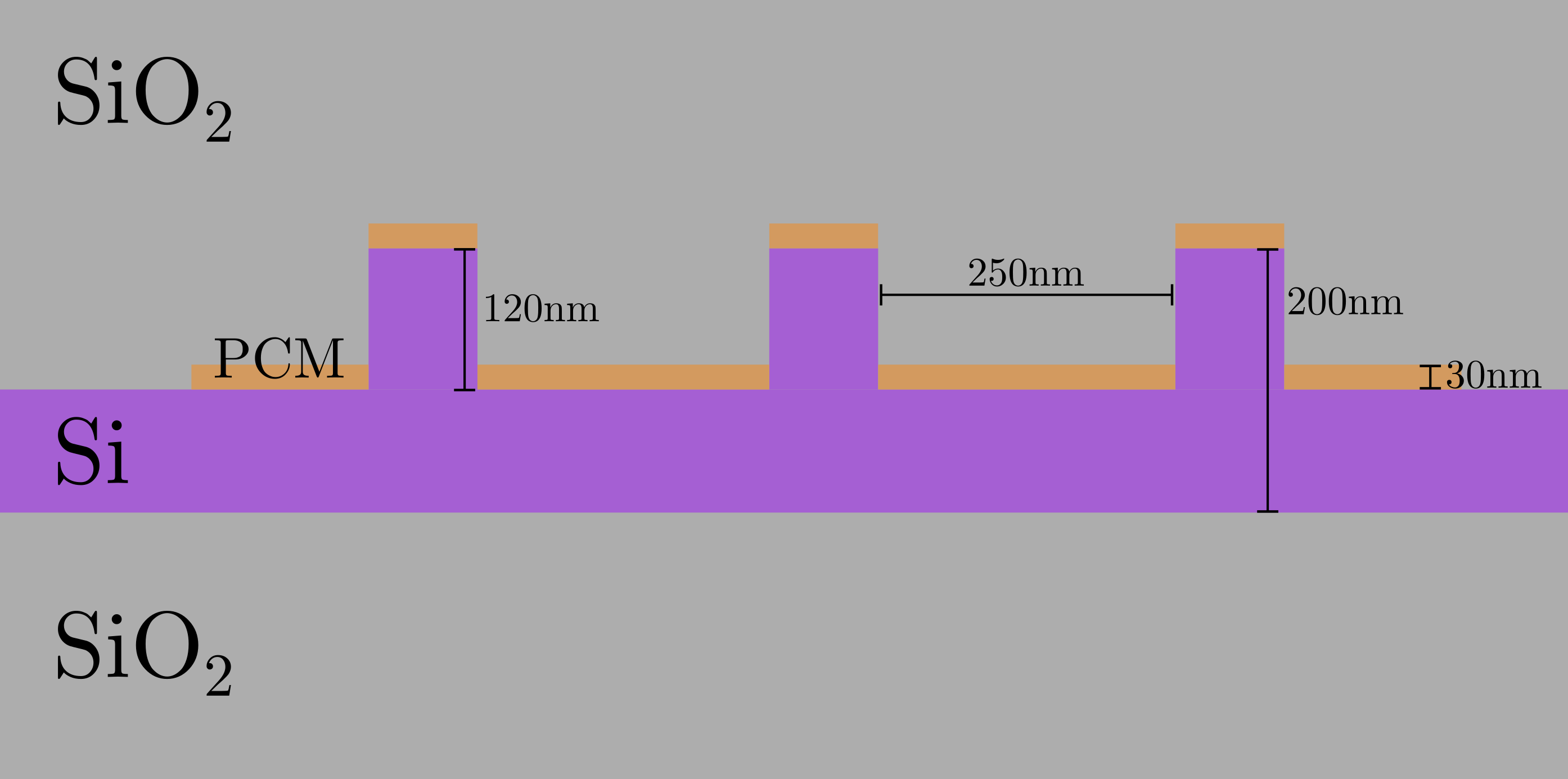}
    \caption{Cross section schematic of the $3\times3$ Si waveguide array inside the coupling reigon. Outside of this waveguides fan out to a spacing of 1\textmu m to avoid cross-talk.}
    \label{Device_cross_section}
\end{figure}


\begin{figure}[htb]
    \centering
    \includegraphics[width = 0.8\linewidth]{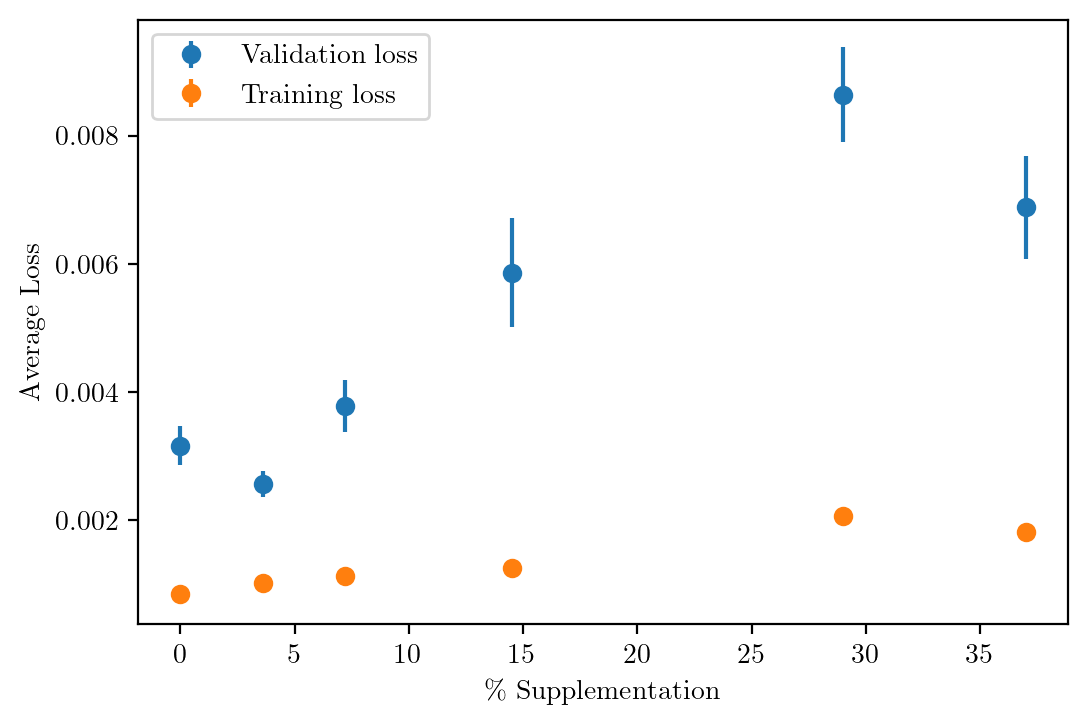}
    \caption{Effect of supplementing a small amount of randomly generated ''noisy'' data. Un-optimized patterns are simulated and added into the initial training data set. Here we present average training losses for the forward network, as well as losses when predicting the transmission matrix for a separate validation dataset. Results are the average of 5 full training cycles with the standard deviation represented by the error bars.}
    \label{fig:SI_supp_data_percentage_network_losses}
\end{figure}

\end{document}